\documentclass[preprint,proceedings]{rmaa}
\shorttitle{Wind velocity at 200 mb}
\shortauthor{Carrasco \& Sarazin}

\SetYear{2003}
\SetConfTitle{San Pedro M\'artir: Astronomical Site Evaluation}  

\title{\bf High altitude wind velocity at  San Pedro M\'artir and Mauna Kea }

\author{Esperanza Carrasco \altaffilmark{1}  \& Marc Sarazin \altaffilmark{2}}

\altaffiltext{1}{Instituto Nacional de  Astrof\'{\i}sica, \'Optica y Electr\'onica,\\ 
Luis Enrique Erro 1, Tonantzintla, Puebla 72840, M\'exico(\protect\email{bec@inaoep.mx}).}

\altaffiltext{2}{European Southern Observatory\\ Karl-Schwarzchild-Str. 2\\ D-85748 Garching, Germany
(\protect\email{msarazin@eso.org}).}

\suppressfulladdresses

\listofauthors{E. Carrasco, M. Sarazin}
\indexauthor{Carrasco, E., Sarazin, M.}
\addkeyword{Site-testing, atmospheric effects}
\resumen{Analizamos el promedio mensual de la velocidad del viento a una altitud aproximada
 de 12 km sobre el nivel del mar, en el periodo comprendido entre 1980 y 1995 en San Pedro M\'artir,
 Mauna Kea, en otros observatorios y en algunos sitios de inter\'es.  Comparamos los resultados
 obtenidos de dos bases de datos, GGUAS y NCEP. Los resultados muestran que San Pedro M\'artir y Mauna
 Kea son comparables y se encuentran entre los lugares mas adecuados para aplicar t\'ecnicas
 que compensan las deformaciones del frente de onda que cambian lentamente.}

\abstract{We analyze the   monthly  average wind velocity at  about 12 km  above sea  level,
between 1980 to 1995, for San Pedro M\'artir,  Mauna Kea, another existing observatories and some
  sites of interest. We compare the results obtained from two different 
 data sets,  the GGUAS and NCEP.  Our results show that San Pedro M\'artir
 and Mauna Kea are comparable and are amongst  the most suitable sites
to apply  slow wavefront corrugation correction techniques.}
 
\begin{document}
\maketitle 
\listofauthors{E. Carrasco \& M. Sarazin}

\indexauthor{Carrasco, E.}
\indexauthor{Sarazin, M. }

     One of the most important parameters  
  of  the potential sites for extremely 
  large telescopes projects, is their suitability  for adaptive optics. 
  Sarazin \& Tokovinin  (2002) have shown that such a suitability 
 is related to the low velocity of upper atmospheric motion,  at about 12 km above
 sea level.  Here, we analyse the  wind velocity  monthly variation over a period of 16 years,
 for San Pedro M\'artir (SPM),  Mauna Kea  and  some  main observatory sites.

  The  National Oceanic and  Atmospheric Administration (NOAA) in the USA runs
  several climatological projects.  Within the NOAA, the  National Climatic 
  Data Center (NCDC) is in charge of  managing the resource of global 
  climatological in-situ and remotely sensed data and  information. Weather data
  from the atmosphere  are obtained from instrument packages such as radiosondes
  and  rawinsondes  carried   by weather balloons that 
  transmit the data back to a receiving station on the ground. The upper air 
  data consists of temperature, relative humidity, atmospheric pressure, and wind.

  In a similar way the NOAA Climate Diagnostics Center (CDC) goal is to identify
the nature and causes for climate variations on time scales ranging from a month to
centuries. The CDC NCEP/NCAR   (National Center for Environmental Prediction/National Center
for Atmospheric Research) Global Reanalysis Project is using the state-of-the-art 
analysis and forecast system to perform data assimilation using past data from 1948 
to the present. The  NCEP's role is to use a current and fixed (Jan. 1995) 
version of a data assimilation and operational forecast model.
The  task of NCAR is to collect and organize many of the land and marine 
surface data archives, to provide these to NCEP along with many observed upper and aircraft 
observations, receive and store the output archives.

  The Global Gridded Upper Air Statistics (GGUAS) 1980-1995 Version 1.1 is 
  distributed by the NCDC. The source of the 
  GGUAS data set was the European Centre for Medium-Range Weather Forecasts 
  (ECMWF) 0000Z and 1200Z gridded analyses. The GGUAS data set describes the 
  atmosphere for each month of the year represented on a 2.5 degree global 
  grid at 15 standard pressure levels. We  use the monthly average and
  rms scalar wind velocity at 200 mb (about 12 km above sea level) at the grid
   point closest to the 
  sites on the basis of two records per day.

   The CDC Derived NCEP Reanalysis Products  include over 80 different variables and several
different coordinates systems at 0Z, 6Z, 12Z, and 18Z forecasted values.
In particular,  the  Derived NCEP Pressure Level product provides, the monthly
wind speed on a 2.5 degree global grid at 17 pressure levels.  We are
using the monthly wind velocity  at 200 mb on the basis of four records per day 
 between 1980 and 1995 to compare with the  GGUAS data for the same period.

 The  coordinates of the sites included in this analysis are shown
 in   table \ref{coordinates}.  Two sets of  coordinates
 are given for each site.  The first ones are the actual coordinates  used as
 input to the data basis that correspond to the grid points closest to the 
 geographical coordinates, the latter  ones  are given as a reference.  Costa Rica 
 is included as a tropical place where there is not jet stream.

  Table \ref{gguas} shows the results obtained from the GGUAS data base for
 the sites given in table \ref{coordinates}.  
The appearence in some sites of large discrepancies between the NOAA NCEP 
results and previously published data based on GGUAS statistics (Sarazin, 2000, 
Sarazin, 2002) led to a revision of the latter and the discovery of a bug in the
query script. The GGUAS statistics presented here thus supersede all previously 
published data.

 In  table \ref{ncep}  the  corresponding wind velocities obtained  from the NCEP  
 reanalysis are shown. The first column indicates the month number. For each site
 two values are reported. First, the wind velocity obtained by averaging the monthly values
 over the 16 years period. Second, the instantaneous rms fluctuations around the average. 
For the whole period, the annual average and  the quadratic average of the monthly
rms values are  included.  The latter represents a typical fluctuation of the wind speed velocity.
It must be noticed that the rms fluctuations are larger for the GGUAS data  than for
the NCEP reanalysis data by a factor larger than 2.

 Table \ref{compara} shows a summary of the  wind  velocity   yearly average  obtained
with the GGUAS and the NCEP data sets.  Here we include  the error in the annual average determination,
given by the  annual rms fluctuations divided by $\sqrt N$ where N is equal to 12.

\begin{figure*}[!t]
     \includegraphics[height=17.5cm]{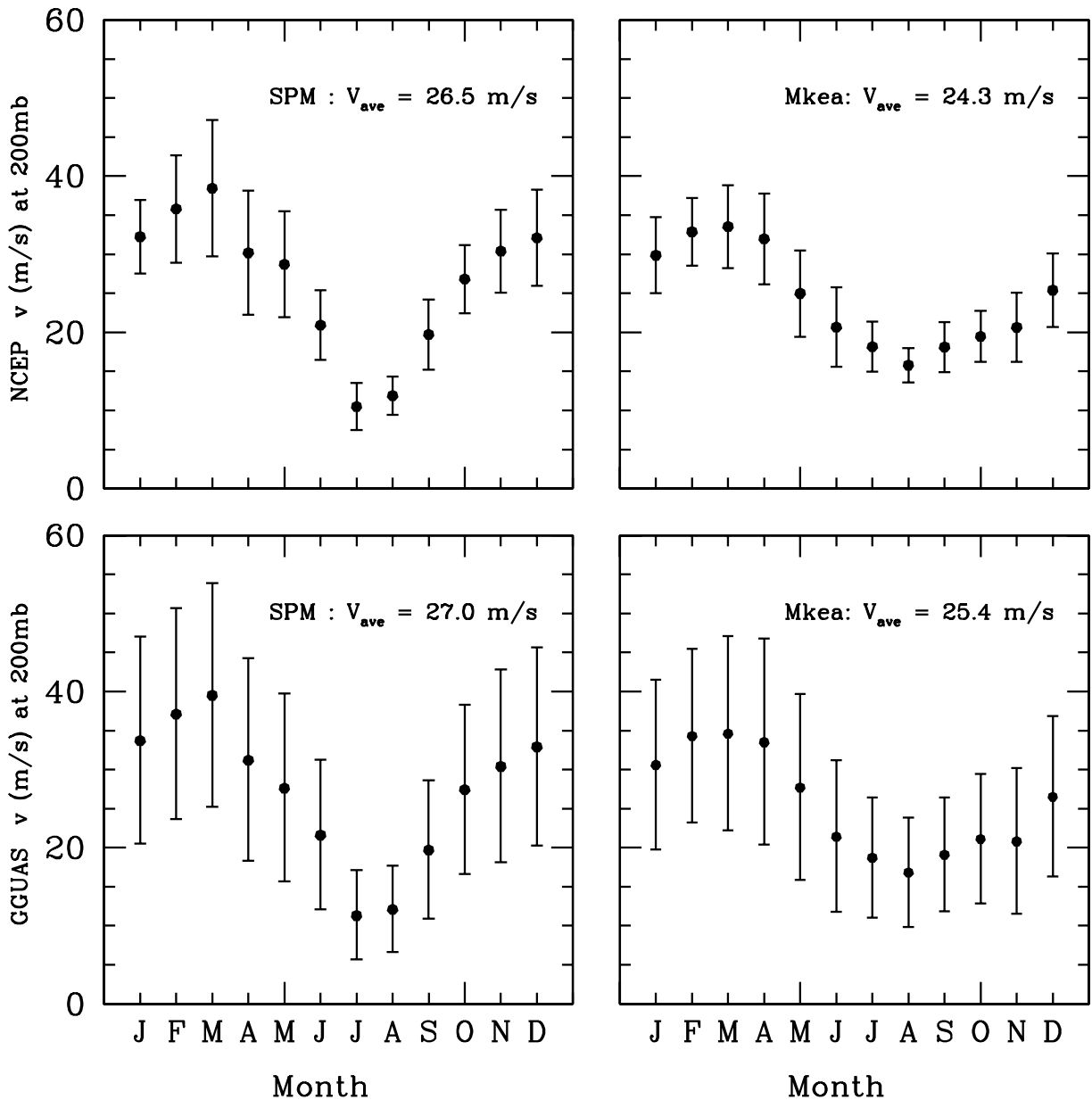}%
  \caption{Monthly average wind speed at 200 mb  for SPM  and
 Mauna Kea. The NCEP reanalysis data for both sites are shown in the
 upper panels. The corresponding  GGUAS data are  shown  in the lower panels.  
 The error bars were calculated assuming a wind log normal distribution
 as it is explained in the text. For SPM the results obtained from the
 GGUAS  and the NCEP data basis, show the same seasonal trend. Similarly
 the results for Mauna Kea have the same seasonal trend in the GGUAS and 
 the NCEP data sets.  However, the instantaneous  wind velocity fluctuations  
  represented by the error bars are less than half for the NCEP
  than for the GGUAS data.  It must be noticed that in July and August the
  monthly average wind speed is lower in SPM than in Hawaii.}

  \label{spm_mkea}
\end{figure*}

   Figure \ref{spm_mkea} is a plot of the monthly average wind speed at 200 mb  for SPM (left) and
 Mauna Kea (right).  In the upper panels the NCEP reanalysis data
 are shown for both sites  with the corresponding  GGUAS data
 in the lower panels.  
 The error bars were calculated assuming a wind log normal distribution:
\begin{equation}
p(v) dv = {1\over b\sqrt{2\pi}} \exp\left\{-{1\over 2}
\left({\ln v - a \over b}\right)^{2} \right\}  {dv\over v} \quad .
\end{equation}

The mean wind speed is here equal to $\bar{v} = e^{a + b^{2}}$ and 
asymmetrical error bars can be constructed from the positions of the 
$\pm 1\sigma$  probability points\footnote {where the cumulative probabilities are
p=0.843 and p=0.1587}, 
$\bar{v} - \sigma_{-} = e^{a - b}$ and $\bar{v} + \sigma_{-} = e^{a + b}$.

   The  monthly average wind velocity obtained with the GGUAS and NCEP reanalysis data sets 
show the same  seasonal trend for SPM and Mauna Kea. Nevertheless,  the  instantaneous
fluctuations are larger, more than a factor of two, for the GGUAS than for the NCEP data.  
   In fact,  the results in table \ref{compara} show that with the GGUAS data, 
the yearly average wind velocities  obtained for all the sites are comparable within 3$\sigma$. 
If we consider the results from the NCEP reanalysis for the sites with the extremes values,
La Silla and La Palma  the difference in the  annual  wind velocity
is larger that 5.5$\sigma$. Therefore to compare sites is more accurate to use  the results 
obtained from the NCEP reanalysis data set.

     The  wind velocity annual average for SPM is  1.2$\sigma$, 1.6$\sigma$, 2.6$\sigma$ below
Maidanak,  Paranal and  La Silla respectively. On the other hand, SPM is 1$\sigma$, 1.2$\sigma$, 
 2.1$\sigma$  above Mauna Kea,  Gamsberg and   La Palma respectively.
For Mauna Kea, the annual average wind velocity is 2.4$\sigma$, 2.6$\sigma$, 4.1$\sigma$ below
Maidanak,  Paranal and  La Silla respectively. In contrast, Mauna Kea is 1.6$\sigma$, 1.8$\sigma$
above Gamsberg and La Palma respectively,  giving a  statistically tangible  site ranking. 
Nevertheless, a deeper and fairer  comparison  would requiere a  monthly
based data analysis. For instance, the monthly average wind speed at SPM in July and August
is lower than at Mauna Kea and La Palma.

We have analysed  the seasonal variations of the  monthly average wind velocity
over a  16 year period for some of the main observatory sites in the 
world by using data from  GGUAS and the NCEP reanalysis data sets. We conclude that
the data obtained from the NCEP reanalysis are more accurate for the determination
of the   monthly average wind velocity  at 200 mb.  Using these data we  have
shown that SPM and Mauna Kea are amongst the best observatory sites  suitable for
Adaptive  Optics techniques.

\acknowledgements

{NCEP Reanalysis data provided by  the NOAA-CIRES Climate Diagnostics Center, Boulder, Colorado,
USA, from their Web site at http://www.cdc.noaa.gov/.}

\begin{table*}[!t]\centering
 \newcommand{\DS} {\hspace{1\tabcolsep}}
   \tablecols{28}  
   \caption{ Sites Coordinates }
   \begin{tabular}{l @{\DS} cccc c ccc }
 \toprule
& \multicolumn{2}{c}{ Costa Rica}
    &\multicolumn{2}{c}{SPM} &\multicolumn{2}{c}{Mauna Kea}&\multicolumn{2}{c}{Paranal }\\
     
& \multicolumn{1}{c} {Closest} &\multicolumn{1}{c}{real}& \multicolumn{1}{c} {Closest} & \multicolumn{1}{c}{real}
& \multicolumn{1}{c} {Closest} &\multicolumn{1}{c}{real}& \multicolumn{1}{c} {Closest} &\multicolumn{1}{c}{real}\\
\midrule

Latitud   & $+10.0$  & $+10.0$ & $+30.00 $ & $31.04$ & $+20.00$   & $+19.83$  & $-25.00$  & $-24.63$  \\ 

Longitude & $-85.0$  & $-85.0$ & $-115.00$ & $-115.46$ & $-155.00$  & $-155.47$ & $-70.00$  & $-70.40$  \\
               
 \bottomrule
&\multicolumn{2}{c}{La Silla}&\multicolumn{2}{c}{La Palma}&\multicolumn{2}{c}{Gamsberg} &\multicolumn{2}{c}{Maidanak}\\    
& \multicolumn{1}{c} {Closest} &\multicolumn{1}{c}{real}& \multicolumn{1}{c} {Closest} &\multicolumn{1}{c}{real} 
& \multicolumn{1}{c} {Closest} &\multicolumn{1}{c}{real}& \multicolumn{1}{c} {Closest} &\multicolumn{1}{c}{real}\\

\midrule      
Latitute  &  $-30.00$   & $-29.25$ & $+30.00$ & $+28.76$ & $-22.50$ & $-23.34$ & $+40.00$ & $+38.68$\\

Longitude &  $-70.00$   & $-70.73$ & $-17.50$ & $-17.88$ & $+15.00$ & $+16.23$ & $+65.00$ & $+66.90$\\

 \bottomrule
  \end{tabular}
  \label{coordinates}
\end{table*}

\begin{table*}[!t]\centering
 \newcommand{\DS} {\hspace{1\tabcolsep}}
   \tablecols{28}  
   \caption{Wind Velocity (m/s) at 200 mb obtained from the NOAA GGUAS data base}
   \begin{tabular}{l @{\DS} lccc c cccc cccc c ccc}
 \toprule
\multicolumn{1}{c}{M} & \multicolumn{2}{c}{ Costa Rica}
    &\multicolumn{2}{c}{SPM} &\multicolumn{2}{c}{Mauna Kea}&\multicolumn{2}{c}{Paranal }&\multicolumn{2}{c}{La Silla}
    &\multicolumn{2}{c}{La Palma}&\multicolumn{2}{c}{Gamsberg} &\multicolumn{2}{c}{Maidanak}\\    
& \multicolumn{1}{c} {Ave} &\multicolumn{1}{c}{rms}& \multicolumn{1}{c} {Ave} &\multicolumn{1}{c}{rms}& \multicolumn{1}{c} {Ave} &\multicolumn{1}{c}{rms}
& \multicolumn{1}{c} {Ave} &\multicolumn{1}{c}{rms}& \multicolumn{1}{c} {Ave} &\multicolumn{1}{c}{rms}& \multicolumn{1}{c} {Ave} &\multicolumn{1}{c}{rms}
& \multicolumn{1}{c} {Ave} &\multicolumn{1}{c}{rms}& \multicolumn{1}{c} {Ave} &\multicolumn{1}{c}{rms}\\

\midrule  
         1  & 14.2 & 7.1 & 33.7 & 14.7 & 30.6 & 11.8 &  19.5&7.7     & 27.2&10.0  & 20.9&10.2  & 13.6&6.7    & 28.6&10.2\\
         2  & 14.1 & 6.9 & 37.1 & 14.7 & 34.3 & 11.9 &  19.3&8.3     & 24.6&10.4  & 24.0&10.9  & 12.2&6.5    & 27.9&12.5 \\
         3  & 12.5 & 6.8 & 39.5 & 15.6 & 34.6 & 13.5 &  22.0&8.7     & 26.4&11.3  & 25.7&13.8  & 18.5&8.4    & 27.2&10.6 \\
         4  & 10.8 & 5.9 & 31.2 & 14.6 & 33.5 & 14.6 &  29.7&10.7    & 31.7&12.8  & 29.2&13.4  & 28.7&12.5   & 24.9&10.0 \\
         5  & 7.8  & 4.3 & 27.6 & 13.7 & 27.7 & 13.5 &  35.5&14.5    & 36.2&14.3  & 27.9&12.7  & 30.2&12.0   & 26.5&10.6 \\
         6  &  8.1 & 4.4 & 21.6 & 11.0 & 21.4 & 11.2 &  35.6&14.6    & 36.0&14.1  & 22.9&10.3  & 33.0&12.4   & 30.8&12.3 \\
         7  &  7.6 & 4.0 & 11.3 &  6.9 & 18.7 & 8.6  &  37.4&14.8    & 37.7&15.6  & 16.2&8.6   & 32.1&11.8   & 27.3&9.7   \\
         8  &  8.1 & 4.2 & 12.1 & 6.4  & 16.8 & 7.9  &  36.2&12.6    & 38.1&14.4  & 15.7&8.1    & 29.9&11.8  & 28.9&10.4  \\
         9  &  7.6 & 4.0 & 19.7 & 10.2 & 19.1 & 8.0  &  36.6&13.8    & 36.3&13.8  & 18.2&9.1   & 25.6&10.7   & 30.2&11.0  \\
         10 &  7.8 & 4.2 & 27.4 & 12.0 & 21.1 & 9.2  &  35.8&11.3    & 39.2&13.1  & 19.5&10.6  & 26.6&10.4   & 25.0&11.9 \\
         11 &  8.3 & 4.4 & 30.4 & 13.8 & 20.8 & 10.7 &  30.9&10.0    & 34.0&12.8  & 23.9&11.0  & 24.0&9.4    & 28.2&11.1 \\
         12 & 12.2 & 5.7 & 32.9 &14.0 & 26.5  & 11.4 &  24.5&9.9     & 27.6&11.7  & 21.3&11.4  & 20.9&9.5    & 27.3&11.3 \\
\midrule
Ave & {\bf 9.9} & 5.3  & {\bf  27.0}&12.6& {\bf 25.4} &  11.2   & {\bf  30.3} & 11.7   & {\bf 32.9} &12.9  & {\bf 22.1}&10.9 & {\bf 24.6} &10.4   & {\bf  27.7}&11.0\\       
   
 \bottomrule
  \end{tabular}
  \label{gguas}
\end{table*}

\begin{table*}[!t]\centering
 \newcommand{\DS} {\hspace{1\tabcolsep}}
   \tablecols{18}  
   \caption{Wind Velocity (m/s) at 200 mb obtained from the NOAA NCEP data base}
   \begin{tabular}{l @{\DS} lcccccccccccc c ccc}
 \toprule     
\multicolumn{1}{c}{M} & \multicolumn{2}{c}{ Costa Rica}
    &\multicolumn{2}{c}{SPM} &\multicolumn{2}{c}{Mauna Kea}&\multicolumn{2}{c}{Paranal}&\multicolumn{2}{c}{La Silla}&\multicolumn{2}{c}{La Palma}&\multicolumn{2}{c}{Gamsberg}
    &\multicolumn{2}{c}{Maidanak}\\
    & \multicolumn{1}{c} {Ave} &\multicolumn{1}{c}{rms}& \multicolumn{1}{c} {Ave} &\multicolumn{1}{c}{rms}& \multicolumn{1}{c} {Ave} &\multicolumn{1}{c}{rms}
& \multicolumn{1}{c} {Ave} &\multicolumn{1}{c}{rms}& \multicolumn{1}{c} {Ave} &\multicolumn{1}{c}{rms}& \multicolumn{1}{c} {Ave} &\multicolumn{1}{c}{rms}
& \multicolumn{1}{c} {Ave} &\multicolumn{1}{c}{rms}& \multicolumn{1}{c} {Ave} &\multicolumn{1}{c}{rms}\\    
  
\midrule  
1 & 13.9 &  3.1 &  32.2   &  4.8   & 29.9 &  4.9 &   18.5 &  2.8 & 27.2  &  4.3 & 20.3 &  4.0   & 12.4  &  2.8 &  33.2  &  5.9  \\           
2 & 13.2 &  2.7 &  35.8   &  7.1   & 32.9 &  4.4 &   18.2 &  3.5  & 24.1  &  4.1 & 24.1  &  4.4   & 10.2  &  2.9 &  33.1  &  7.4 \\  
3 & 12.0 &  3.5 &  38.4   &  9.0   & 33.5 &  5.4 &   20.6 &  3.5  & 25.9  &  4.7 & 25.9   &  6.2   & 17.5  &  3.5 &  31.1  &  4.3 \\ 
4 &  9.4 &  2.9 &  30.2   &  8.3   & 32.0 &  5.9 &   28.3 &  3.2 & 30.6  &  4.8 & 29.0  &  4.8  & 27.3  &  4.6 &  26.5  &  4.6 \\ 
5 &  7.1 &  1.3 &  28.7   &  7.1   & 25.0 &  5.7 &   33.9 &  5.3 & 35.6  &  5.2 & 27.4  &  5.4   & 29.3 &  2.9 &  29.3  &  5.3 \\  
6 &  8.7 &  2.1 &  20.9   &  4.6   & 20.7 &  5.3 &   36.5 &  6.2 & 35.1  &  5.7 & 21.4   &  4.2  & 31.7  &  4.3 &  31.5  &  4.4\\ 
7 &  8.4 &  2.0 &  10.5   &  3.2   & 18.2 &  3.3 &   36.7 &  6.3 & 36.9  &  5.6  & 15.7 &  3.9   & 31.5  &  4.6 &  22.2  &  5.4 \\ 
8 &  9.9 &  2.3 &  11.9   &  2.5   & 15.8 &  2.2 &   35.6 &  5.6  & 37.4  &  7.0 & 14.9  &  2.5   & 28.8  &  4.4 &  23.6  &  6.5  \\  
9 &  8.9 &  1.7 &  19.7   &  4.6   & 18.1 &  3.3 &   35.9 &  7.2 & 34.8  &  4.7 & 17.1  &  2.7   & 24.6  &  2.9 &  29.4  &  4.8 \\  
10&  9.1 &  1.8 &  26.8   &  4.4   & 19.5 &  3.3 &   34.9 &  3.2  & 38.2  &  4.1 & 18.81  &  3.8   & 25.0  &  2.8 &  27.4  &  5.0 \\  
11&  8.4 &  2.4 &  30.4   &  5.4   & 20.6 &  4.6 &   29.2 &  3.5 & 34.0  &  6.3 & 21.8  &  4.2   & 22.2  &  3.7 &  31.0  &  4.2  \\ 
12& 12.0 &  2.7
&  32.1   &  6.3   & 25.4 &  4.8 &   23.5 &  3.4  & 34.0  &  6.3 & 20.2  &  5.3   & 18.9  &  2.8 &  31.3  &  5.1 \\ 
\midrule
Ave & {\bf 10.1} &2.4  & {\bf  26.5}   &  5.9 & {\bf 24.3} &  4.5 & {\bf 29.3} &  4.7 & {\bf 32.4} &  5.2 & {\bf 21.4} &  4.4 & {\bf 23.3}  &  3.6 &  {\bf 29.1} &  5.3\\ 
\bottomrule
\end{tabular}
\label{ncep}
\end{table*}

\begin{table*}[!t]\centering
 \newcommand{\DS} {\hspace{1\tabcolsep}}
   \tablecols{4} 
   \caption{Wind velocity annual average }
   \begin{tabular}{l @{\DS} ccc}
 \toprule     

 \multicolumn{1}{c}{Site}& \multicolumn{1}{c}{GGUAS}& \multicolumn{1}{c}{NCEP}\\
                         & \multicolumn{1}{c} { (m/s)} &\multicolumn{1}{c}{ (m/s)} \\
                                       
  \midrule
              SPM        &  {\bf  27.0} $\pm3.6$    & {\bf 26.5}   $\pm 1.7$\\
              Mauna Kea  &  {\bf 25.4}  $\pm 3.2$   & {\bf 24.3}   $\pm 1.3$\\
              Paranal    & {\bf  30.3}  $\pm 3.4$   & {\bf 29.3}   $\pm 1.4$\\
             La Silla    & {\bf 32.9}   $\pm 3.7$   & {\bf 32.4}   $\pm 1.5$\\
             La Palma    & {\bf 22.1}   $\pm 3.1$   & {\bf 21.4}   $\pm 1.3$\\
             Gamsberg    & {\bf 24.6}   $\pm 3.0$   & {\bf 23.3}   $\pm 1.0$\\
             Maidanak    & {\bf 27.7}   $\pm3.2$    & {\bf 29.1}   $\pm 1.5$\\
\bottomrule
\end{tabular}
\label{compara}
\end{table*}

\end{document}